


\def\unlock{
 \catcode`@=11 }
\unlock
\def\lock{\catcode`@=12 }

\newskip\@@@@@
\newskip\@@@@@two
\newskip\t@mp


 \font\twentyfourrm=cmr10               scaled\magstep4
 \font\seventeenrm=cmr10                scaled\magstep3
 \font\fourteenrm=cmr10                 scaled\magstep2
 \font\twelverm=cmr10                   scaled\magstep1
 \font\ninerm=cmr9            \font\sixrm=cmr6
 \font\twentyfourbf=cmbx10              scaled\magstep4
 \font\seventeenbf=cmbx10               scaled\magstep3
 \font\fourteenbf=cmbx10                scaled\magstep2
 \font\twelvebf=cmbx10                  scaled\magstep1
 \font\ninebf=cmbx9            \font\sixbf=cmbx5
 \font\twentyfouri=cmmi10 scaled\magstep4  \skewchar\twentyfouri='177
 \font\seventeeni=cmmi10  scaled\magstep3  \skewchar\seventeeni='177
 \font\fourteeni=cmmi10   scaled\magstep2  \skewchar\fourteeni='177
 \font\twelvei=cmmi10     scaled\magstep1  \skewchar\twelvei='177
 \font\ninei=cmmi9                         \skewchar\ninei='177
 \font\sixi=cmmi6                          \skewchar\sixi='177
 \font\twentyfoursy=cmsy10 scaled\magstep4 \skewchar\twentyfoursy='60
 \font\seventeensy=cmsy10  scaled\magstep3 \skewchar\seventeensy='60
 \font\fourteensy=cmsy10   scaled\magstep2 \skewchar\fourteensy='60
 \font\twelvesy=cmsy10     scaled\magstep1 \skewchar\twelvesy='60
 \font\ninesy=cmsy9                        \skewchar\ninesy='60
 \font\sixsy=cmsy6                         \skewchar\sixsy='60
 \font\twentyfourex=cmex10  scaled\magstep4
 \font\seventeenex=cmex10   scaled\magstep3
 \font\fourteenex=cmex10    scaled\magstep2
 \font\twelveex=cmex10      scaled\magstep1
 \font\elevenex=cmex10      scaled\magstephalf
 \font\twentyfoursl=cmsl10  scaled\magstep4
 \font\seventeensl=cmsl10   scaled\magstep3
 \font\fourteensl=cmsl10    scaled\magstep2
 \font\twelvesl=cmsl10      scaled\magstep1
 \font\ninesl=cmsl9
 \font\twentyfourit=cmti10 scaled\magstep4
 
 \font\fourteenit=cmti10   scaled\magstep2
 \font\twelveit=cmti10     scaled\magstep1
 \font\twentyfourtt=cmtt10 scaled\magstep4
 \font\twelvett=cmtt10     scaled\magstep1
 \font\twentyfourcp=cmcsc10 scaled\magstep4
 \font\twelvecp=cmcsc10    scaled\magstep1
 \font\tencp=cmcsc10
 \font\hepbig=cmdunh10     scaled\magstep5
 \font\hep=cmdunh10        scaled\magstep0
 \newfam\cpfam
 \font\tenfib=cmr10  
 \newcount\f@ntkey            \f@ntkey=0
 \def\samef@nt{\relax \ifcase\f@ntkey \rm \or\oldstyle \or\or
          \or\it \or\sl \or\bf \or\tt \or\caps \fi }

 \def\twentyfourpoint{\relax
     \textfont0=\twentyfourrm           \scriptfont0=\seventeenrm
                                        \scriptscriptfont0=\fourteenrm
     \def\rm{\fam0 \twentyfourrm \f@ntkey=0 }\relax
     \textfont1=\twentyfouri            \scriptfont1=\seventeeni
                                        \scriptscriptfont1=\fourteeni
      \def\oldstyle{\fam1 \twentyfouri\f@ntkey=1 }\relax
     \textfont2=\twentyfoursy           \scriptfont2=\seventeensy
                                        \scriptscriptfont2=\fourteensy
     \textfont3=\twentyfourex           \scriptfont3=\seventeenex
                                        \scriptscriptfont3=\fourteenex
     \def\it{\fam\itfam \twentyfourit\f@ntkey=4 }\textfont\itfam=\twentyfourit
     \def\sl{\fam\slfam \twentyfoursl\f@ntkey=5 }\textfont\slfam=\twentyfoursl
                                        \scriptfont\slfam=\seventeensl
     \def\bf{\fam\bffam \twentyfourbf\f@ntkey=6 }\textfont\bffam=\twentyfourbf
                                        \scriptfont\bffam=\seventeenbf
                                        \scriptscriptfont\bffam=\fourteenbf
     \def\tt{\fam\ttfam \twentyfourtt \f@ntkey=7 }\textfont\ttfam=\twentyfourtt
     \def\caps{\fam\cpfam \twentyfourcp \f@ntkey=8 }
                                        \textfont\cpfam=\twentyfourcp
     \setbox\strutbox=\hbox{\vrule height 23pt depth 8pt width\z@}
     \samef@nt}
 \def\fourteenpoint{\relax
     \textfont0=\fourteenrm          \scriptfont0=\tenrm
     \scriptscriptfont0=\sevenrm
      \def\rm{\fam0 \fourteenrm \f@ntkey=0 }\relax
     \textfont1=\fourteeni           \scriptfont1=\teni
     \scriptscriptfont1=\seveni
      \def\oldstyle{\fam1 \fourteeni\f@ntkey=1 }\relax
     \textfont2=\fourteensy          \scriptfont2=\tensy
     \scriptscriptfont2=\sevensy
     \textfont3=\fourteenex          \scriptfont3=\twelveex
     \scriptscriptfont3=\tenex
     \def\it{\fam\itfam \fourteenit\f@ntkey=4 }\textfont\itfam=\fourteenit
     \def\sl{\fam\slfam \fourteensl\f@ntkey=5 }\textfont\slfam=\fourteensl
     \scriptfont\slfam=\tensl
     \def\bf{\fam\bffam \fourteenbf\f@ntkey=6 }\textfont\bffam=\fourteenbf
     \scriptfont\bffam=\tenbf     \scriptscriptfont\bffam=\sevenbf
     \def\tt{\fam\ttfam \twelvett \f@ntkey=7 }\textfont\ttfam=\twelvett
     \def\caps{\fam\cpfam \twelvecp \f@ntkey=8 }\textfont\cpfam=\twelvecp
     \setbox\strutbox=\hbox{\vrule height 12pt depth 5pt width\z@}
     \samef@nt}

 \def\twelvepoint{\relax
     \textfont0=\twelverm          \scriptfont0=\ninerm
     \scriptscriptfont0=\sixrm
      \def\rm{\fam0 \twelverm \f@ntkey=0 }\relax
     \textfont1=\twelvei           \scriptfont1=\ninei
     \scriptscriptfont1=\sixi
      \def\oldstyle{\fam1 \twelvei\f@ntkey=1 }\relax
     \textfont2=\twelvesy          \scriptfont2=\ninesy
     \scriptscriptfont2=\sixsy
     \textfont3=\twelveex          \scriptfont3=\elevenex
     \scriptscriptfont3=\tenex
     \def\it{\fam\itfam \twelveit \f@ntkey=4 }\textfont\itfam=\twelveit
     \def\sl{\fam\slfam \twelvesl \f@ntkey=5 }\textfont\slfam=\twelvesl
     \scriptfont\slfam=\ninesl
     \def\bf{\fam\bffam \twelvebf \f@ntkey=6 }\textfont\bffam=\twelvebf
     \scriptfont\bffam=\ninebf     \scriptscriptfont\bffam=\sixbf
     \def\tt{\fam\ttfam \twelvett \f@ntkey=7 }\textfont\ttfam=\twelvett
     \def\caps{\fam\cpfam \twelvecp \f@ntkey=8 }\textfont\cpfam=\twelvecp
     \setbox\strutbox=\hbox{\vrule height 10pt depth 4pt width\z@}
     \samef@nt}

 \def\tenpoint{\relax
     \textfont0=\tenrm          \scriptfont0=\sevenrm
     \scriptscriptfont0=\fiverm
     \def\rm{\fam0 \tenrm \f@ntkey=0 }\relax
     \textfont1=\teni           \scriptfont1=\seveni
     \scriptscriptfont1=\fivei
     \def\oldstyle{\fam1 \teni \f@ntkey=1 }\relax
     \textfont2=\tensy          \scriptfont2=\sevensy
     \scriptscriptfont2=\fivesy
     \textfont3=\tenex          \scriptfont3=\tenex
     \scriptscriptfont3=\tenex
     \def\it{\fam\itfam \tenit \f@ntkey=4 }\textfont\itfam=\tenit
     \def\sl{\fam\slfam \tensl \f@ntkey=5 }\textfont\slfam=\tensl
     \def\bf{\fam\bffam \tenbf \f@ntkey=6 }\textfont\bffam=\tenbf
     \scriptfont\bffam=\sevenbf     \scriptscriptfont\bffam=\fivebf
     \def\tt{\fam\ttfam \tentt \f@ntkey=7 }\textfont\ttfam=\tentt
     \def\caps{\fam\cpfam \tencp \f@ntkey=8 }\textfont\cpfam=\tencp
     \setbox\strutbox=\hbox{\vrule height 8.5pt depth 3.5pt width\z@}
     \samef@nt}



 \newtoks\date
 \def\monthname{\relax\ifcase\month 0/\or January\or February\or
    March\or April\or May\or June\or July\or August\or September\or
    October\or November\or December\else\number\month/\fi}
 \date={October 6, 1992}
 \def\today{\the\day\ \monthname\ \the\year}

 \def\\{\relax\ifmmode\backslash\else$\backslash$\fi}
 \def\nextline{\unskip\nobreak\hskip\parfillskip\break}
 
 \def\journal#1&#2,#3(#4){\unskip \enskip {\sl #1}~{\bf #2}, {\rm #3 (19#4)}}
 \def\cropen#1{\crcr\noalign{\vskip #1}}
 
 \def\topspace{\hrule height 0pt depth 0pt \vskip}
\def\nullbox#1#2#3{\vbox to #1
     {\vss\vtop to#2
       {\vss\hbox to #3 {}}}}
\def\n@ll{\nullbox{5pt}{3pt}{2pt}}    
\def\UNRTXT#1{\vtop{\hbox{#1}\kern 1pt \hrule}}
\def\undertext#1{\ifvmode\ifinner \UNRTXT{#1}
                         \else    $\hbox{\UNRTXT{#1}}$
                         \fi
                 \else   \ifmmode \hbox{\UNRTXT{#1}}
                         \else    \UNRTXT{#1}
                         \fi
                 \fi }



\def\low#1{\kern-0.11em\lower0.4em\hbox{$\scriptstyle #1 $}\hskip-0.08em}
\def\up#1{\kern-0.45em\raise0.4em\hbox{$\scriptstyle #1 $}\hskip0.29em}


 \let\int=\intop         
 \def\prop{\mathrel{{\mathchoice{\pr@p\scriptstyle}{\pr@p\scriptstyle}{
                 \pr@p\scriptscriptstyle}{\pr@p\scriptscriptstyle} }}}
 \def\pr@p#1{\setbox0=\hbox{$\cal #1 \char'103$}
    \hbox{$\cal #1 \char'117$\kern-.4\wd0\box0}}
 \let\sec@nt=\sec
 \def\sec{\relax\ifmmode\let\n@xt=\sec@nt\else\let\n@xt\section\fi\n@xt}
 \def\@versim#1#2{\lower0.2ex\vbox{\baselineskip\z@skip\lineskip\z@skip
   \lineskiplimit\z@\ialign{$\m@th#1\hfil##\hfil$\crcr#2\crcr\sim\crcr}}}
\def\lsim{\mathrel{\mathpalette\@versim<}}
\def\gsim{\mathrel{\mathpalette\@versim>}}



\def\part#1#2{\partial^{#1}\kern-.1667em\hbox{$#2$}}
\def\der#1#2{d^{#1}\kern-.1067em\hbox{$#2$}}

\def\delt#1{{\delta\kern-.15em #1 }}



\def\rightbracearrow{\raise0.5em\hbox{$|$}\mkern-8.0mu\rightarrow}


\def\U#1{U\kern-.25em \left( #1 \right) }




 \normalbaselineskip = 20pt plus 0.2pt minus 0.1pt
 \normallineskip = 1.5pt plus 0.1pt minus 0.1pt
 \normallineskiplimit = 1.5pt
 \newskip\normaldisplayskip
    \normaldisplayskip = 20pt plus 5pt minus 10pt
 \newskip\normaldispshortskip
    \normaldispshortskip = 6pt plus 5pt
 \newskip\normalparskip
    \normalparskip = 6pt plus 2pt minus 1pt
 \newskip\skipregister
    \skipregister = 5pt plus 2pt minus 1.5pt
 \newif\ifsingl@    \newif\ifdoubl@
 \newif\ifb@@kspace   \b@@kspacefalse
 \newif\iftwelv@    \twelv@true
 \newif\ifp@genum


 \def\singlespace{\singl@true\doubl@false\b@@kspacefalse\spaces@t}
 \def\doublespace{\singl@false\doubl@true\b@@kspacefalse\spaces@t}
 \def\normalspace{\singl@false\doubl@false\b@@kspacefalse\spaces@t}
 \def\bookspace{\b@@kspacetrue\singl@false\doubl@false\spaces@t}
 \def\tablespace{\singlespace}
 \def\smashspace{\singl@false\doubl@false\b@@kspacefalse\subspaces@t3:8;}


 \def\Tenpoint{\tenpoint\twelv@false\spaces@t}
 \def\Twelvepoint{\twelvepoint\twelv@true\spaces@t}


 \def\spaces@t{\relax
       \iftwelv@ \ifsingl@\subspaces@t3:4;\else\subspaces@t1:1;\fi
        \else \ifsingl@\subspaces@t1:2;\else\subspaces@t4:5;\fi \fi
       \ifb@@kspace
          \baselineskip=14pt
       \fi
       \ifdoubl@ \multiply\baselineskip by 5
          \divide\baselineskip by 4 \fi }
 \def\subspaces@t#1:#2;{
       \baselineskip = \normalbaselineskip
       \multiply\baselineskip by #1 \divide\baselineskip by #2
       \lineskip = \normallineskip
       \multiply\lineskip by #1 \divide\lineskip by #2
       \lineskiplimit = \normallineskiplimit
       \multiply\lineskiplimit by #1 \divide\lineskiplimit by #2
       \parskip = \normalparskip
       \multiply\parskip by #1 \divide\parskip by #2
       \abovedisplayskip = \normaldisplayskip
       \multiply\abovedisplayskip by #1 \divide\abovedisplayskip by #2
       \belowdisplayskip = \abovedisplayskip
       \abovedisplayshortskip = \normaldispshortskip
       \multiply\abovedisplayshortskip by #1
         \divide\abovedisplayshortskip by #2
       \belowdisplayshortskip = \abovedisplayshortskip
       \advance\belowdisplayshortskip by \belowdisplayskip
       \divide\belowdisplayshortskip by 2
       \smallskipamount = \skipregister
       \multiply\smallskipamount by #1 \divide\smallskipamount by #2
       \medskipamount = \smallskipamount \multiply\medskipamount by 2
       \bigskipamount = \smallskipamount \multiply\bigskipamount by 4 }
 \def\normalbaselines{ \baselineskip=\normalbaselineskip
    \lineskip=\normallineskip \lineskiplimit=\normallineskip
    \iftwelv@\else \multiply\baselineskip by 4 \divide\baselineskip by 5
      \multiply\lineskiplimit by 4 \divide\lineskiplimit by 5
      \multiply\lineskip by 4 \divide\lineskip by 5 \fi }



 \newskip\tablelineskip           \tablelineskip=0.7in
 \newskip\figurelineskip          \figurelineskip=1.55in
 \newskip\tablelinelength     \tablelinelength=4.5in
 \newskip\tabledotvskip       \tabledotvskip=-0.359in
 \newskip\tabledothskip       \tabledothskip=-0.306in%
 \newskip\bookchapterskip     \bookchapterskip=1.0in
 \newcount\chapternumber    \chapternumber=0
 \newcount\appendixnumber   \appendixnumber=0
 \newcount\sectionnumber    \sectionnumber=0
 \newcount\subsectionnumber \subsectionnumber=0
 \newcount\equanumber       \equanumber=1
 \newcount\problemnumber    \problemnumber=0
 \newcount\figurecount      \figurecount=1
 \newcount\conpage          \conpage=0
 \let\chapterlabel=0
 \newtoks\constyle          \constyle={\Number}
 \newtoks\appendixstyle     \global\appendixstyle={\Alphabetic}
 \newtoks\chapterstyle      \chapterstyle={\Number}
 \newtoks\subsecstyle       \subsecstyle={\alphabetic}
 \newskip\chapterskip       \chapterskip=\bigskipamount
 \newskip\sectionskip       \sectionskip=\medskipamount
 \newskip\headskip          \headskip=8pt plus 3pt minus 3pt
 \newif\ifsp@cecheck
 \newif\iffirst@ppendix       \global\first@ppendixtrue
 \newdimen\chapterminspace    \chapterminspace=15pc
 \newdimen\sectionminspace    \sectionminspace=8pc
 \interlinepenalty=50
 \interfootnotelinepenalty=5000
 \predisplaypenalty=9000
 \postdisplaypenalty=500
 \newwrite\tableconwrite
 \newbox\tableconbox
 \newif\iftabl@conlist
 \newwrite\figwrite
 \newwrite\figurewrite
 \newif\iffigur@list
 \newwrite\eqnwrite
 \newif\if@qlist
 \newif\ifeqlo@d            \eqlo@dfalse
 \newwrite\tablewrite
 \newwrite\tableswrite
 \newif\iftabl@list
 \newwrite\appendixwrite
 \newif\if@ppendix
 \newcount\@ppcharnumber     \@ppcharnumber=64
 \newcount\referencecount \newbox\referencebox
 \newcount\tablecount
 \newif\ifindex      \indexfalse
 \newwrite\indexwrite
 \newbox\indexbox
 \newskip\indexskip
 \newif\ifmanualpageno   \manualpagenofalse
 \newdimen\letterhsize       \letterhsize=6.5in
 \newdimen\lettervsize       \lettervsize=8.5in
 \newdimen\labelhsize        \labelhsize=8.5in
 \newdimen\labelvsize        \labelvsize=11.0in
 \newdimen\letterhoffset     \letterhoffset=0.0in
 \newdimen\lettervoffset     \lettervoffset=0.250in


 \def\Number#1{\number #1}
 \def\makel@bel{\xdef\chapterlabel{%
     \the\chapterstyle{\the\chapternumber}.}}
 \def\sectionlabel{\number\sectionnumber }
 \def\subseclabel{{\the\subsecstyle{\the\subsectionnumber}. }}
 \def\alphabetic#1{\count255='140 \advance\count255 by #1\char\count255}
 \def\Alphabetic#1{\count255=64 \advance\count255 by #1\char\count255}
 \def\Roman#1{\uppercase\expandafter{\romannumeral #1}}


 \countdef\pagenumber=1  \pagenumber=1
 \def\advancepageno{\global\advance\pageno by 1
    \ifnum\pagenumber<0 \global\advance\pagenumber by -1
     \else\global\advance\pagenumber by 1 \fi \global\frontpagefalse }
\def\pagefolio#1{\ifnum#1<0 \romannumeral-#1
            \else \number#1 \fi }

\def\folio{\pagefolio{\pagenumber}}
 \def\pagecontents{
    \ifvoid\topins\else\unvbox\topins\vskip\skip\topins\fi
    \dimen@ = \dp255 \unvbox255
    \ifvoid\footins\else\vskip\skip\footins\footrule\unvbox\footins\fi
    \ifr@ggedbottom \kern-\dimen@ \vfil \fi }
 \def\makeheadline{\vbox to 0pt{ \hfuzz=30pt \skip@=\topskip
       \advance\skip@ by -12pt \advance\skip@ by -2\normalbaselineskip
       \vskip\skip@ \line{\vbox to 12pt{}\the\headline\hfill} \vss
       }\nointerlineskip}
 \def\makefootline{\baselineskip = 1.5\normalbaselineskip
                  \line{\the\footline}}
 \def\nopagenumbers{\p@genumfalse}
 \def\pagenumbers{\p@genumtrue}


 \def\tableconbreakfill{\hfill\break}


 \let\conbreak=\tableconbreak



 \def\splitprep{\global\newlinechar=`\^^J}
 \def\splitprepend{\global\newlinechar=-1}

 \def\consection#1{
    \iftabl@conlist
       \splitprep
       \immediate\write\tableconwrite{\string\immediate
                                      \string\spacecheck\sectionminspace
                                     }
       \immediate\write\tableconwrite{\string\vskip0.25in
                         \string\titlestyle{\string\bf\ #1 }%
                         \string\vskip0.0in
                         \string\nullbox{1pt}{1pt}{1pt}%
                                     }
       \splitprepend
    \fi}
 \def\unnumberedchapters{\let\makel@bel=\relax \let\chapterlabel=\relax
             \let\sectionlabel=\relax \let\subseclabel=\relax \equanumber=-1 }
 \def\titlestyle#1{\par\begingroup \interlinepenalty=9999
      \leftskip=0.02\hsize plus 0.23\hsize minus 0.02\hsize
      \rightskip=\leftskip \parfillskip=0pt
      \hyphenpenalty=9000 \exhyphenpenalty=9000
      \tolerance=9999 \pretolerance=9000
      \spaceskip=0.333em \xspaceskip=0.5em
      \iftwelv@\fourteenpoint\else\twelvepoint\fi
    \noindent #1\par\endgroup }
 \def\spacecheck#1{\p@gecheck{#1}%
    \ifsp@cecheck
       \else \vfill\eject
    \fi}
 \def\majorreset{
    \ifnum\figurecount<0\else\global\figurecount=1\fi
    \ifnum\equanumber<0 \else\global\equanumber=1\fi
    \sectionnumber=0 \subsectionnumber=0
    \tablecount=0  \problemnumber=0
    \bookheadline={}%
    \chapterheadline={}%
    }
 \def\chapterreset{\global\advance\chapternumber by 1
                   \majorreset
                   \makel@bel}
 \def\appendflag#1{
        \if@ppendix
        \else
          \global\@ppendixtrue
          \starttable{APPENDICES CALLED}{\appendixout}{\appendixwrite}{6}
        \fi
        \@ddconentry{\appendixwrite}{\noindent{\bf #1}}{1}
                   }


 \def\ack{\par\penalty-100\medskip \spacecheck\sectionminspace
      \ifIEEE\bf\noindent{Acknowledgements}\rm%
       \else\line{\fourteenrm\hfil ACKNOWLEDGEMENTS\hfil}\fi%
      \nobreak\vskip\headskip }


 \def\Textindent#1{\noindent\llap{#1\enspace}\ignorespaces}


 \newtoks\temphold
 \newtoks\ta
 \newtoks\tb
 \newtoks\captiontoks
 \newwrite\capwrite
 \newcount\runner  \runner=0
 \newcount\wordsnum \wordsnum=0
\newif\iflongc@p  \longc@pfalse
\def\longcap{\bgroup\obeyspaces\endlinechar=-1
             \global\longc@ptrue}
\def\endlongcap{\egroup\global\longc@pfalse}

 \def\confolio{\pagefolio{\conpage}}
 \def\p@gecheck#1{
        \dimen@=\pagegoal
        \advance\dimen@ by -\pagetotal
        \ifdim\dimen@<#1
           \global\sp@cecheckfalse
           \else\global\sp@cechecktrue
        \fi}
 \def\st@rtt@ble#1#2#3{
                 \message{listing #1:  type
                          \noexpand#2 for list}
                 \global\setbox#3=\vbox{\normalbaselines
                      \titlestyle{\seventeenrm\bf #1} \vskip\headskip}}
 \def\starttable#1#2#3#4{%
     \message{external listing of #1:  type %
              \noexpand#2 for list}%
     \ifcase#4%
           \immediate\openout#3=TABLE_OF_CONTENTS.TEX
       \or%
           \immediate\openout#3=FIG_CAP_AND_PAGE.TEX
       \or%
           \immediate\openout#3=FIGURE_CAPTIONS.TEX
       \or%
           \immediate\openout#3=TABLE_CAPTIONS.TEX
       \or%
           \immediate\openout#3=TAB_CAP_AND_PAGE.TEX
       \or%
           \immediate\openout#3=EQN_PAGE.TEX
       \or%
           \immediate\openout#3=APP_CALLED.TEX
       \or%
           \immediate\openout#3=REFERENC.TEX
      \else \immediate\message{You are in big trouble call a %
                                 TeXnician}%
     \fi%
     \ifIEEE
        \ifnum#4=7\immediate\write#3{\string\vbox {\string\normalbaselines%
              \string\bf\noindent References \string\vskip \string\headskip}}
        \fi%
     \else\immediate\write#3{\string\vbox {\string\normalbaselines%
               \string\titlestyle {\string\seventeenrm \string\bf \ #1}
               \string\vskip \string\headskip}}\fi%
 }%

\def\untoken#1#2{
     \ta=\expandafter{#1}\tb=\expandafter{#2}%
     \immediate\edef\temphold{\the\ta\the\tb}
     \global\captiontoks=\expandafter{\temphold}}%

\def\getc@pwrite#1{
                   \ifx#1\endcaption%
                      \let\next=\relax%
                      \immediate\write\capwrite{\the\captiontoks}
                      \global\captiontoks={}
                    \else%
                       \untoken{\the\captiontoks}{#1}
                       \ifx#1\space%
                             \advance\runner by 1
                             \advance\wordsnum by1 \let\next=\getc@pwrite
                       \fi%
                    \let\next=\getc@pwrite\fi%
                 \ifnum\runner=10
                       \immediate\message{.}
                       \immediate\write\capwrite{\the\captiontoks}
                       \global\captiontoks={}
                       \runner=0
                 \fi
                       \next}%

\def\c@pwrite#1#2{
               \let\capwrite=#1
               \global\runner=0
               \immediate\message{working on long caption. .}
               \wordsnum=0 \getc@pwrite#2\endcaption
               \immediate\message{The number of words= \number\wordsnum}}

\def\t@@bl@build@r#1#2{\relax
         \let\conbreak=\tableconbreakfill 
         \conpage=#2
         \t@mp=\tablelineskip
         \multiply\t@mp by #1
         \@@@@@=\tablelinelength \advance\@@@@@ by -\t@mp%
                                 \advance\@@@@@ by -1in%
             \@@@@@two=\tablelinelength%
             \advance\@@@@@two by -1in%
             \vglue\separationskip
         \vbox\bgroup
             \vbox\bgroup\parshape=3 0pt\@@@@@two%
                               \t@mp\@@@@@%
                               \t@mp\@@@@@%
                   \parfillskip=0pt%
                   \pretolerance=9000 \tolerance=9999 \hbadness=2000%
                   \hyphenpenalty=9000 \exhyphenpenalty=9000%
                   \interlinepenalty=9999%
                   \tablespace
                   \vskip\baselineskip\raggedright\noindent}
\def\endt@@bl@build@r{{\tabledotfill \hskip-0.07in $\,$}%
                  \egroup%
             \parshape=0%
            \vskip\tabledotvskip\hskip\@@@@@two\hskip\tabledothskip%
            { \tabledotfill\quad} {\confolio}%
         \egroup
         \vskip-0.10in}%

 \def\addconentry#1#2{%
     \setbox0=\vbox{\normalbaselines #2}%
     \relax%
     \global\setbox#1=\vbox{\unvbox #1 \vskip 4pt plus 2pt minus 1pt \box0 }}%

\def\@ddconentry#1#2#3{
     \conpage=\pagenumber%
     \p@gecheck{0pt}%
     \ifsp@cecheck \else
         \if\conpage<0
               \advance\conpage by -1
            \else
               \advance\conpage by 1
         \fi%
     \fi%
     \let\conbreak=\tableconbreakfill
     \splitprep
     \immediate\write#1
              {\string\t@@bl@build@r}%
     \immediate\write#1{{#3}%
                        {\the\conpage}%
                       }%
     \iflongc@p
         \immediate\c@pwrite{#1}{ #2 }%
       \else
         \immediate\write#1{{#2}}%
     \fi
     \immediate\write#1
              {\string\endt@@bl@build@r}%
     \splitprepend
}

\def\@ddfigure#1#2#3#4{
        \splitprep
        \captiontoks={}
        \immediate\write#1{{\string\vglue\string\separationskip}%
                          }%
        \immediate\write#1
              {\string\par \string\nullbox {1pt}{1pt}{1pt}%
              }%
        \immediate\write#1{\string\vbox}%
        \immediate\write#1{\string\bgroup \string\tablespace}
        \immediate\write#1{\string\parshape=2}
        \immediate\write#1{0pt\string\tablelinelength}
        \immediate\write#1{\string\figurelineskip\string\@@@@@}
        \immediate\write#1{\string\bgroup
                           }
       \iflongc@p
            \immediate\c@pwrite{#1}{#4~#2:\quad\ #3}
          \else
            \immediate\write#1{#4~#2:\quad\ #3}
       \fi
       \immediate\write#1{\string\egroup
                          \string\egroup
                         }%
       \immediate\write#1{\string\parshape=0%
                          }%
       \splitprepend
                     }%

 \def\captionsetup{
        \@@@@@=\tablelinelength \advance\@@@@@ by -\figurelineskip}
 \def\manualpageno{\manualpagenotrue}
 
 \def\dumplist#1{%
     \bookheadline={}%
     \chapterheadline={}%
     \unlock   
     \global\let\conbreak=\tableconbreakfill
     \ifmanualpageno
         \else
            \pagenumber=1
     \fi
     \ifb@@kstyle
        \global\multiply\pagenumber by -1
        \bookheadline={}
     \fi
     \vskip\chapterskip  
     \ifcase#1
            \chapterheadline={Contents}
            \bookheadline={Contents}    
            \input TABLE_OF_CONTENTS.TEX
       \or
            \chapterheadline={Contents}
            \bookheadline={Contents}         
            \input User_table_of_contents.tex
       \or
            \chapterheadline={Contents}
            \bookheadline={Contents}         
            \input User_table_of_contents.tex
            \input TABLE_OF_CONTENTS.TEX
       \or
           {\obeylines
            \input index.tex
           }
       \or
           {\obeylines
            \input User_index.tex
           }
       \or
           {\obeylines
            \input User_index.tex
            \input index.tex%
           }
       \or
           \captionsetup
           \input FIGURE_CAPTIONS.TEX
       \or
           \input FIG_CAP_AND_PAGE.TEX
       \or
           \captionsetup
           \input TABLE_CAPTIONS.TEX
       \or
           \input TAB_CAP_AND_PAGE.TEX
       \or
           \input EQN_PAGE.TEX
       \or
           \input APP_CALLED.TEX
       \or
           \input REFERENC.TEX
       \else \immediate\message{ Call a TeXnician, You are in BIG trouble}
     \fi
     \vfill\eject 
     \lock}
 \def\dumpbox#1{         
     \ifb@@kstyle
        \multiply\pagenumber by -1
        \bookheadline={}
     \fi
     \vskip\chapterskip  
     \unvbox#1           
     \vfill\eject}       


 \def\notabledots{\global\let\tabledotfill=\hfill}
 \def\tabledots{\global\let\tabledotfill=\dotfill}
 \def\tableconlist{\global\tabl@conlisttrue
                   \starttable{CONTENTS}{\conout}{\tableconwrite}
                              {0}}
 \def\tableconlistoff{\global\tabl@conlistfalse
                      \immediate\closeout\tableconwrite}

 \def\reflist{\global\referencelisttrue}

 \def\figurelist{\global\figur@listtrue
                 \starttable{FIGURE CAPTIONS AND PAGES}
                       {\figuresout}{\figurewrite}{1}
                 \starttable{FIGURE CAPTIONS}{\figout}{\figwrite}{2}
                }
 \def\figurelistoff{\global\figur@listfalse}

 \def\tablelist{\global\tabl@listtrue
                \starttable{TABLE CAPTIONS}{\tabout}{\tablewrite}{3}
                \starttable{TABLE CAPTIONS AND PAGES}
                           {\tablesout}{\tableswrite}{4}
               }
 \def\tablelistoff{\global\tabl@listfalse}


 \def\produceequations#1{
     \ifnum\equanumber<0
          \immediate\write\equ@tions{\string\xdef \string#1
                  {{\string\rm  (\number-\equanumber)}} }
       \else
          \immediate\write\equ@tions{\string\xdef \string#1
                  {{\string\rm  (\chapterlabel\number\equanumber)}} }
     \fi}


 \def\equ@tionlo@d{ \ifeqlo@d
                       \else \input BOOK_EQUATIONS
                             \eqlo@dtrue
                       \fi}


 \def\appendixout{\immediate\closeout\appendixwrite
                  \dumplist{11}
                 }
 \def\conout{\normalspace\immediate\closeout\tableconwrite
             \dumpbox{\tableconbox}
             \dumplist{0}}

 \def\refout{\immediate\closeout\referencewrite
             \dumplist{12}
   }
 \def\figout{
             \immediate\closeout\figwrite
             \dumplist{6}}
 \def\figuresout{
             \normalspace\immediate\closeout\figurewrite
             \dumplist{7}}
 
 \def\tabout{\immediate\closeout\tablewrite
             \dumplist{8}}
 \def\tablesout{\normalspace\immediate\closeout\tableswrite
          \dumplist{9}}
 
 \def\eqout{\immediate\closeout\eqnwrite
          \dumplist{10}
          \immediate\closeout\equ@tions}



\newdimen\referenceminspace  \referenceminspace=25pc
\newcount\referencecount     \referencecount=0
\newcount\titlereferencecount     \titlereferencecount=0
\newcount\lastrefsbegincount \lastrefsbegincount=0
\newdimen\refindent     \refindent=30pt
\newif\ifreferencelist       \global\referencelisttrue
\newif\ifreferenceopen       \global\referenceopenfalse
\newwrite\referencewrite
\newwrite\equ@tions
\newtoks\rw@toks


\def\NPrefmark#1{\attach{\scriptscriptstyle [ #1 ] }}
\def\PRrefmark#1{\attach#1}
\def\IEEErefmark#1{ [#1]}
\def\refmark#1{\relax%
     \ifPhysRev\PRrefmark{#1}%
     \else%
        \ifIEEE\IEEErefmark{#1}%
          \else%
            \NPrefmark{#1}%
          \fi%
     \fi}


\def\REF#1#2{\space@ver{}\refch@ck
   \global\advance\referencecount by 1 \xdef#1{\the\referencecount}%
   \r@fwrit@{#1}{#2}}%
\def\ref#1{\REF\?{#1}\refend}%
\def\Ref#1#2{\REF#1{#2}\refend}%
\def\refend{\global\lastrefsbegincount=\referencecount
            \refsend}
\def\REFS#1#2{\REF#1{#2}\global\lastrefsbegincount=\referencecount\relax}%
\def\REFSCON{\REF}%
\def\refsend{\refmark{\count255=\referencecount 
   \advance\count255 by-\lastrefsbegincount %
   \ifcase\count255 %
       \number\referencecount
   \or%
      \number\lastrefsbegincount,\number\referencecount
   \else%
      \number\lastrefsbegincount -\number\referencecount
  \fi}}%


\def\refch@ck{\ifreferencelist%
                  \ifreferenceopen%
                       \else \global\referenceopentrue%
                           \starttable{REFERENCES}{\refout}{\referencewrite}{7}
                  \fi%
              \fi%
}


\def\rw@begin#1\splitout{\rw@toks={#1}\relax%
   \immediate\write\referencewrite{\the\rw@toks}\futurelet\n@xt\rw@next}%
\def\rw@next{\ifx\n@xt\rw@end \let\n@xt=\relax%
      \else \let\n@xt=\rw@begin \fi \n@xt}%
\let\rw@end=\relax%
\let\splitout=\relax%
\def\r@fwrit@#1#2{%
   \splitprep%
   \immediate\write\referencewrite{\ifIEEE\noexpand\refitem{\noindent[#1]}
                                   \else\noexpand\refitem{#1.}\fi}%
   \rw@begin #2\splitout\rw@end \@sf%
   \splitprepend%
                 }%


\def\refitem#1{\par \hangafter=0 \hangindent=\refindent \Textindent{#1}}%



%
%


\def\TITLEREF#1#2{\space@ver{}\refch@ck%
   \global\advance\titlereferencecount by 1%
   \xdef#1{\alphabetic{\the\titlereferencecount}}%
   \r@fwrit@{#1}{#2}}%
%



 \def\checkreferror{\ifinner\errmessage{This is a wrong place to DEFINE
     a reference or a figure! You may have your references / figure
     captions screwed up. }\fi}


 \def\FIG#1#2{\checkreferror
     \ifnum\figurecount<0
          \global\xdef#1{\number-\figurecount}%
          \global\advance\figurecount by -1
       \else
          \global\xdef#1{\chapterlabel\number\figurecount}%
          \global\advance\figurecount by 1
     \fi
     \iffigur@list
        \@@@@@=\hsize \advance\@@@@@ by -1.219in
        \@ddconentry{\figurewrite}%
            {\noindent
                \bgroup \bf \expandafter{\csname\string#1.\endcsname} \egroup
                \quad{#1}\quad\ #2
            }
            {1}%
        \@ddfigure{\figwrite}{#1}{#2}{Figure}%
     \fi     }%



 \def\nexttable{\checkreferror\global\advance\tablecount by 1}


 \def\TABLE#1#2{\nexttable\xdef#1{\chapterlabel\the\tablecount}%
     \iftabl@list
        \@ddfigure{\tablewrite}{#1}{#2}{Table}%
        \@ddconentry{\tableswrite}%
               {\noindent
                   \bgroup \bf \expandafter{\csname\string#1.\endcsname}
\egroup
                   \quad{#1}\quad\ #2}
               {1}%
     \fi}%



 \def\eqname#1{\relax
     \if@qlist
       \produceequations{#1}
     \fi
     \ifnum\equanumber<0
           \global\xdef#1{{\rm(\number-\equanumber)}}
           \global\advance\equanumber by -1
     \else
           \global\xdef#1{{\rm(\chapterlabel \number\equanumber)}}
           \global\advance\equanumber by 1
     \fi
     \if@qlist
       \@ddconentry{\eqnwrite}
           {\noindent
             \bgroup \bf \expandafter{\csname\string#1\endcsname.} \egroup
             \quad{#1}\quad}
           {1}
     \fi      }



 \def\eqinsert#1{\noalign{\dimen@=\prevdepth \nointerlineskip
    \setbox0=\hbox to\displaywidth{\hfil #1}
    \vbox to 0pt{\vss\hbox{$\!\box0\!$}\kern-0.5\baselineskip}
    \prevdepth=\dimen@}}



 \def\globaleqnumbers{\relax\if\equanumber<0\else\global\equanumber=-1\fi}


\def\globalfigurenumbers{\relax\if\figurecount<0\else\global\figurecount=-1\fi}



 \newskip\lettertopfil      \lettertopfil = 0pt plus 1.5in minus 0pt
 \newskip\spskip            \setbox0\hbox{\ } \spskip=-1\wd0
 \newskip\signatureskip     \signatureskip=40pt
 \newskip\letterbottomfil   \letterbottomfil = 0pt plus 2.3in minus 0pt
 \newskip\frontpageskip     \frontpageskip=1\medskipamount plus .5fil
 \newskip\headboxwidth      \headboxwidth= 0.0pt
 \newskip\letternameskip    \letternameskip=0pt
 \newskip\letterpush        \letterpush=0pt
 \newbox\headbox
 \newbox\headboxbox
 \newbox\physbox
 \newbox\letterb@x
 \newif\ifse@l              \se@ltrue
 \newif\iffrontpage
 \newif\ifletterstyle
 \newif\ifhe@dboxset        \he@dboxsetfalse
 \newtoks\memoheadline
 \newtoks\letterheadline
 \newtoks\letterfrontheadline
 \newtoks\lettermainheadline
 \newtoks\letterfootline
 \newdimen\holder
 \newdimen\headboxwidth
 \newtoks\myletterheadline  \myletterheadline={define \\myletterheadline}
 \newtoks\mylettername    \mylettername={{\twentyfourbf define}\
\\mylettername}
 \newtoks\phonenumber       \phonenumber={\rm (313) 764-4437}
 \newtoks\faxnumber	    \faxnumber={(313) 763-9694}
 \newtoks\telexnumber       \telexnumber={\tenfib 4320815 UOFM UI}


 \def\FIRSTP@GE{\ifvoid255\else\vfill\penalty-2000\fi
                \global\frontpagetrue}
 \def\FRONTPAGE{\ifvoid255\else\vfill\penalty-2000\fi
       \masterreset\global\frontpagetrue
       \global\lastp@g@no=-1 \global\footsymbolcount=0}
 
 \letterfootline={\hfil}
 \lettermainheadline={\hbox to \hsize{
        \rm\ifp@genum page \ \folio\fi
        \hfill\today}}
 \letterheadline{\hfuzz=60pt\iffrontpage\the\letterfrontheadline
      \else\the\lettermainheadline\fi}
 \def\addressee#1{\vskip-0.188in \singlespace
    \ialign to\hsize{\strut ##\hfill\tabskip 0pt plus \hsize \cr #1\hfill\crcr}
    \medskip\noindent\hskip\spskip}
 \def\letterstyle{\global\letterstyletrue%
                  \global\paperstylefalse%
                  \global\b@@kstylefalse%
                  \global\frontpagetrue%
                  \global\singlespace\global\lettersize}
 \def\lettersize{\hsize=\letterhsize \vsize=\lettervsize
                 \hoffset=\letterhoffset \voffset=\lettervoffset
    \headboxwidth=\hsize  \advance\headboxwidth by -1.05in
    \skip\footins=\smallskipamount \multiply\skip\footins by 3}
 \def\telex#1{\telexnumber={\tenfib #1}}
 \def\noseal{\se@lfalse}


 \def\myaddress#1{%
         \singlespace
         \global\he@dboxsettrue
         {\hsize=\headboxwidth \setbox0=\vbox{\ialign{##\hfill\cr #1 }}
         \global\setbox\headboxbox=\vbox {\hfuzz=50pt
                  \vfill
                  \line{\hfil\box0\hfil}}}}
 \def\MICH@t#1{\hfuzz=50pt\global\setbox\physbox=\hbox{\it#1}\hfuzz=1pt}
 \def\l@tt@rcheck{\ifhe@dboxset\else
                     \immediate\message{Setting default letterhead.}
                     \immediate\message{For more information consult the PHYZZM
                                       documents.}
                     \UM
                  \fi}
 \def\l@tt@rs#1{\l@tt@rcheck
              \vfill\supereject 
              \global\letterstyle
              \global\letterfrontheadline={}
              \global\setbox\headbox=\vbox{{
                  \hbox to\headboxwidth{\hepbig\hfil #1\hfil}
                  \vskip 1pc
                  \unvcopy\headboxbox}}}
 \def\HEAD#1{\singlespace  
    \global\letternameskip=22pt plus 0pt minus 0pt
    \global\memoheadline={UM\ #1\ Memorandum}
    \mylettername={\hep #1}
       \myaddress{
   \hep\hfill\hbox{The Harrison M. Randall Laboratory of Physics}\hfill \cr
   \hep\hfill\hbox{500 East University, Ann Arbor, Michigan 48109-1120}\hfill
\cr
                                                 \cr}}


 \def\theory{\phonenumber={\rm (313) 763-9698}
        \HEAD{Theoretical Physics}
        \PUBNUM{TH} \MICH@t{Theoretical Physics}}
 \def\UM{\phonenumber={\rm (313) 764-4437}
        \HEAD{Department of Physics}
        \PUBNUM{PHYSICS} \MICH@t{Department of Physics}}
 

 \def\lettertext{\par\unvcopy\letterb@x\par}
 \def\multiletter{\setbox\letterb@x=\vbox\bgroup
       \everypar{\vrule height 1\baselineskip depth 0pt width 0pt }
       \singlespace \topskip=\baselineskip }
 \def\letterend{\par\egroup}

 \def\letters{\l@tt@rs{The University of Michigan}}
 \def\letter{\wlog{\string\letter}
        \vfill\supereject
        \normalspace       
        \FRONTPAGE
        \nullbox{1pt}{1pt}{1pt}\vskip-0.8750in
        \setbox0 = \vbox{\hfuzz=50pt   
           {\hsize=\headboxwidth
	   \unvcopy\headbox\hfill
           \singlespace
           \vskip-0.1in
	   \dimen1=2.25truein
	   \setbox1=\hbox{\hep(313) 763-4929}     \advance\dimen1 by \wd1
	   \setbox1=\hbox{\hep\the\phonenumber}   \advance\dimen1 by -1\wd1
           \centerline{\the\mylettername\hskip\dimen1{\hep\the\phonenumber}}}}
%
%
 \ifse@l
   \dimen0=0.8in
   \setbox1=\hbox to \dimen0{%
    \dimen0=1.0in
    \vbox to \dimen0{
      \vss
        \UGbody{tex$inputs:umseal.ps}{1.0}{8.31}{1}                 
     }%
     \hss
    }
 \fi
        \hfuzz=5pt
        \ifse@l
          \hfuzz=15pt\line{\hbox to0pt{}\hskip 0.3in\box1\box0 \hfill}
	\else
	  \line{\hfill\box0\hfill}
	\fi
        \vskip0.35in
        \vskip\letterpush
        \rightline{\today}
        \addressee}

 \def\myletter{\l@tt@rs{\the\myletterheadline}\letter}

 \def\signed#1{\par \penalty 9000 \bigskip \dt@pfalse
   \everycr={\noalign{\ifdt@p\vskip\signatureskip\global\dt@pfalse\fi}}
   \setbox0=\vbox{\singlespace \halign{\tabskip 0pt \strut ##\hfill\cr
    \noalign{\global\dt@ptrue}#1\hfill\crcr}}
   \line{\hskip 0.5\hsize minus 0.5\hsize \box0\hfill} \medskip }
 \def\copies#1{\singlespace\hfill\break
   \line{\nullbox{0pt}{0pt}{\hsize}}
   \setbox0 = \vbox {
     \noindent{\tenrm cc:}\hfill 
     \vskip-0.2115in\hskip0.000in\vbox{\advance\hsize by-\parindent
       \ialign to\hsize{\strut ##\hfill\hfill
                 \tabskip 0pt plus \hsize \cr #1\crcr}}
       \hbox spread\hsize{}\hfill\vfill}
   \line{\box0\hfill}}
 \def\endletter{\nullbox{0pt}{0pt}{\hsize}
       \ifnum\pagenumber=1
                \vskip\letterbottomfil\vfill\supereject
          \else
                \vfill\supereject
       \fi
       \wlog{ENDLETTER}}



 \newif\ifp@bblock          \p@bblocktrue
 \newif\ifpaperstyle
 \newif\ifb@@kstyle
 \newtoks\paperheadline
 \newtoks\bookheadline
 \newtoks\chapterheadline
 \newtoks\paperfootline
 \newtoks\bookfootline
 \newtoks\Pubnum            \Pubnum={$\caps UM - PHY - PUB -
                                        \the\year - \the\pubnum $}
 \newtoks\pubnum            \pubnum={00}
 \newtoks\pubtype           \pubtype={\tensl Preliminary Version}
 \newcount\yeartest
 \newcount\yearcount

 \def\sequentialfootnotes{\global\seqf@@tstrue}
 \def\PHYSREV{\paperstyle\PhysRevtrue\PH@SR@V
     \let\refmark=\attach}
 \def\PH@SR@V{\doubl@true \baselineskip=24.1pt plus 0.2pt minus 0.1pt
              \parskip= 3pt plus 2pt minus 1pt }
 \def\IEEE{\paperstyle\IEEEtrue\I@EE\doublespace\rm\let\refmark=\IEEErefmark%
             \let\unnumberedchapters=\relax}
 \def\I@EE{\baselineskip=24.1pt plus 0.2pt minus 0.1pt
              \parskip= 3pt plus 2pt minus 1pt }


 \def\bookstyle{\b@@kstyletrue%
                \letterstylefalse%
                \paperstylefalse%
                \equ@tionlo@d
                \Tenpoint
                \frenchspacing
                \parskip=0pt
                \bookspace\booksize}
 \def\paperstyle{\paperstyletrue%
                 \b@@kstylefalse%
                 \letterstylefalse%
                 \normalspace}

\def\booksize{\hsize=29pc\vsize=45pc\hoffset=0.85in\voffset=0.475in\hfuzz=2.5pc
               \itemsize=\parindent}


 \paperfootline={\hss\iffrontpage
                         \else \ifp@genum
                                  \hbox to \hsize{\tenrm\hfill\folio\hfill}\hss
                               \fi
                      \fi}
 \paperheadline={\hfil}
 \bookfootline={\hss\iffrontpage
                       \ifp@genum
                              \hbox to \hsize{\tenrm\bf\hfill\folio\hfill}\hss
                       \fi
                     \else
                           \hbox to \hsize{\hfill\hfill}\hss
                     \fi}


 \def\titlepage{
    \yeartest=\year \advance\yeartest by -1900
    \ifnum\yeartest>\yearcount
          \global\PUBNUM{UM}
    \fi
    \FRONTPAGE\paperstyle\ifPhysRev\PH@SR@V\fi
    \ifp@bblock\p@bblock\fi}

 \def\nopubblock{\p@bblockfalse}
 \def\endpage{\vfil\break}
 \def\p@bblock{\begingroup \tabskip=\hsize minus \hsize
    \baselineskip=1.5\ht\strutbox \topspace-2\baselineskip
    \halign to\hsize{\strut ##\hfil\tabskip=0pt\crcr
    \the\Pubnum\cr \the\date\cr \the\pubtype\cr}\endgroup}


 \def\PUBNUM#1{
     \yearcount=\year
     \advance\yearcount by -1900
     \Pubnum={$\caps UM- #1 - \the\yearcount - \the\pubnum $}}
 \def\title#1{\vskip\frontpageskip \titlestyle{#1} \vskip\headskip }
 \def\author#1{\vskip\frontpageskip\titlestyle{\twelvecp #1}\nobreak}

 \def\address#1{\par\noindent\titlestyle{\twelvepoint\it #1}}

 \def\andaddress{\par\kern 5pt \centerline{\sl and} \address}
 \def\abstract{\vskip\frontpageskip\centerline{\fourteenrm ABSTRACT}
               \vskip\headskip }




 \newtoks\foottokens       \foottokens={\Tenpoint\singlespace}
 \newdimen\footindent      \footindent=24pt
 \newcount\lastp@g@no	   \lastp@g@no=-1
 \newcount\footsymbolcount \footsymbolcount=0
 \newif\ifPhysRev
 \newif\ifIEEE
 \newif\ifseqf@@ts         \global\seqf@@tsfalse



 \def\footrule{\dimen@=\prevdepth\nointerlineskip
    \vbox to 0pt{\vskip -0.25\baselineskip \hrule width 0.35\hsize \vss}
    \prevdepth=\dimen@ }
 \def\vfootnote#1{\insert\footins\bgroup  \the\foottokens
    \interlinepenalty=\interfootnotelinepenalty \floatingpenalty=20000
    \splittopskip=\ht\strutbox \boxmaxdepth=\dp\strutbox
    \leftskip=\footindent \rightskip=\z@skip
    \parindent=0.5\footindent \parfillskip=0pt plus 1fil
    \spaceskip=\z@skip \xspaceskip=\z@skip
    \Textindent{$ #1 $}\footstrut\futurelet\next\fo@t}


 \def\footnote#1{\attach{#1}\vfootnote{#1}}


 \let\footsymbol=\star
 \def\footsymbolgen
 {  \relax
   \ifseqf@@ts \seqf@@tgen
   \else
      \ifPhysRev
         \iffrontpage \NPsymbolgen
         \else \PRsymbolgen
         \fi
      \else \NPsymbolgen
      \fi
   \fi
   \footsymbol
 }
 \def\seqf@@tgen{\ifnum\footsymbolcount>0 \global\footsymbolcount=0\fi
       \global\advance\footsymbolcount by -1
       \xdef\footsymbol{\number-\footsymbolcount} }
 \def\NPsymbolgen
 {  \ifnum\footsymbolcount<0 \global\footsymbolcount=0\fi
   {  \iffrontpage \relax
      \else
         \ifnum \lastp@g@no = \pageno
            \relax
         \else
            \global\lastp@g@no = \pageno
            \global\footsymbolcount=0
         \fi
      \fi
   }
   \ifcase\footsymbolcount
      \fd@f\star \or \fd@f\dagger \or \fd@f\ast \or
      \fd@f\ddagger \or \fd@f\natural \or \fd@f\diamond \or
      \fd@f\bullet \or \fd@f\nabla
   \fi
   \global\advance\footsymbolcount by 1
   \ifnum\footsymbolcount>6 \global\footsymbolcount=0\fi
 }
 \def\fd@f#1{\xdef\footsymbol{#1}}
 \def\PRsymbolgen{\ifnum\footsymbolcount>0 \global\footsymbolcount=0\fi
       \global\advance\footsymbolcount by -1
       \xdef\footsymbol{\sharp\number-\footsymbolcount} }
 \def\space@ver#1{\let\@sf=\empty \ifmmode #1\else \ifhmode
    \edef\@sf{\spacefactor=\the\spacefactor}\unskip${}#1$\relax\fi\fi}
 \def\attach#1{\space@ver{\strut^{\mkern 2mu #1} }\@sf\ }


 \footline={\ifletterstyle \the\letterfootline
               \else \ifpaperstyle \the\paperfootline
                        \else \the\bookfootline
                      \fi
            \fi}
 \headline={\let\conbreak=\tableconbreakspace%
            \ifletterstyle \the\letterheadline
               \else \ifpaperstyle \the\paperheadline
                        \else \iffrontpage {}
                                  \else
                                     \setbox0=\hbox{\ {\tenrm\bf \folio}}
                                     \advance\hsize by \wd0
                                     \ifodd\pagenumber
                                          \hbox to \hsize{%
                                              \the\bookheadline\hfill\hfill
                                              \box0
                                                         }%
                                        \else
                                          \hskip-\wd0
                                          \hbox to \hsize{%
                                              \box0\hfill\hfill
                                              \the\chapterheadline
                                              }%
                                     \fi
                              \fi
                     \fi
             \fi}



 \def\masterreset{\global\pagenumber=1 \global\chapternumber=0
    \global\appendixnumber=0
    \global\equanumber=1 \global\sectionnumber=0 \global\subsectionnumber=0
    \global\referencecount=0 \global\figurecount=1 \global\tablecount=0
    \global\problemnumber=0
    \global\@ppendixfalse
    \ifIEEE\global\setbox\referencebox=\vbox{\normalbaselines
          \noindent{\bf References}\vskip\headskip}%
     \else\global\setbox\referencebox=\vbox{\normalbaselines
          \centerline{\fourteenrm REFERENCES}\vskip\headskip}\fi
   }




 \normalspace                      
 \masterreset
 \pagenumbers                      
 \Twelvepoint                      
 \figurelistoff                    
 \tableconlistoff                  
 \tablelistoff                     
 \reflist                          
 \telex{4320815 UOFM UI}           
 \headboxwidth=6.5in
 \advance\headboxwidth by -1.05in  
 \UM                               
 \paperstyle                       
 \tabledots                        
 \manualpageno                     
 \he@dboxsetfalse                  
 \lock
 \hfuzz=1pt
 \vfuzz=0.2pt

\message{PHYZZM version 1.9}

\def\hcirc{{h^\circ}}
\def\ss{\scriptscriptstyle}
\def\zss{{\scriptscriptstyle Z}}
\def\un{\underbar}
\input tables
\theory
\pubnum{24}
\pubtype{}
\titlepage
\title{CALCULABLE UPPER LIMIT ON THE MASS OF THE LIGHTEST HIGGS
BOSON IN ANY PERTURBATIVELY VALID SUPERSYMMETRIC THEORY}
\author{G.L. Kane, Chris Kolda, and James D. Wells}
\address{Randall Physics Laboratory\nextline
University of Michigan\nextline
Ann Arbor, MI   48109-1120}
\abstract
We show that there is a calculable upper limit on the mass of the
lightest Higgs boson in any supersymmetric theory that remains
perturbative up to a high scale .  There are no restrictions on the
Higgs sector, or the gauge group or particle content.  We estimate the
value of the upper limit to be $m_{\hcirc} < 146$ GeV for 100 GeV
$\lsim M_t \lsim 145$ GeV, from all effects except possibly additional
heavy fermions beyond top (which could increase the limit by 0-20 GeV
if any existed); for $M_t \gsim$ 145 GeV the limit decreases
monotonically.  We expect to be able to decrease the value of the upper
limit by at least a few percent by very careful analysis of the
conditions.  It is not normal in models for the actual mass to
saturate the upper limit.
\endpage
\noindent {\bf INTRODUCTION}

In the minimal supersymmetric Standard Model (MSSM), with two Higgs
$SU(2)$ doublet fields, the fermionic partners of the Higgs bosons
must have gauge couplings.  Then the supersymmetry leads to gauge
couplings also for the Higgs boson self-interactions,
so their masses can be expressed
in terms of vacuum expectation values and gauge couplings.
This is well known to lead to a tree level upper bound
\Ref\inoue{K. Inoue, A. Kakuto, H. Komatsu and S. Takeshita, Prog. Theor.
Phys., \un{67} (1982) 1889, \un{68} (1982) 927; R. Flores and M. Sher,
Ann. Phys. \un{148} (1983) 95.}
$m^2_{\hcirc} < M^2_\zss\cos^2 2\beta,$
where $\tan \beta$ is the ratio of the vacuum expectation values of
the two Higgs fields that give mass to up-type and down-type quarks,
and $\hcirc$ is the lightest Higgs boson.  It is remarkable that this
upper limit is independent of the scale of supersymmetry masses and
the scale of supersymmetry breaking, and holds independently of the
short distance or large mass behavior of the theory.

Radiative corrections shift this limit,\Ref\haber{H. Haber and R.
Hempfling, Phys. Rev. Lett. \un{66} (1991) 1815; Y. Okada, M.
Yamaguchi, and T. Yanagida, Prog. Theor. Phys. \un{85} (1991) 1; Phys.
Lett. \un{B262} (1991) 54; J. Ellis, G. Ridolfi, and F. Zwirner,
Phys. Lett. \un{B257} (1991) 83.}adding to $m_\hcirc$ a numerically important
contribution proportional to $M_t^2$ and logarithmically dependent on
squark masses.

In the minimal (non-supersymmetric) Standard Model (SM) there is also
an upper limit\Ref\cabibbo{N. Cabibbo, L. Maiani, G. Parisi, and R.
Petronzio, Nucl. Phys. \un{B158} (1979) 295.} on the Higgs boson mass
$m_\varphi$ if one adds
the condition that the couplings of the theory should be quantities
that can be calculated perturbatively up to a scale of order $10^{16}$
GeV.  Most people believe that this condition is now implied by
experiment\Ref\lang{P. Langacker and M. Luo, Phys. Rev. \un{D44} (1991) 817;
U. Amaldi, W. de Boer, and H. Furstenau, Phys. Lett. \un{B260} (1991) 447;
see also F. Anselmo, L. Cifarelli, and A. Zichichi, CERN/LAA/MSL/92-011,
to appear in Il Nuovo Cimento.}
since the gauge couplings do
approximately meet at a point, or alternatively, starting with the
symmetry value $\sin^2\theta_{\ss W}=3/8$ at about $10^{16}$
GeV, the calculated value of $\sin^2\theta_{\ss W}=0.23$ at our scale
agrees with experiment in some theories to about 1\% accuracy.
Imposing this condition does not require a belief in any particular
grand unified theory (GUT), or even in a GUT at all, but only that any
acceptable theory should remain perturbative up to a scale of about
$10^{16}$ GeV.  Results are also insensitive to the choice of the GUT
scale.  A phrase is needed to describe such a condition, so we denote
the requirement that any candidate theory should remain perturbative
up to a scale of order $10^{16}$ GeV as ``perturbative validity'',
whether or not unification is assumed.  Anyone who does not require
that acceptable theories be perturbatively valid must maintain that
the above results are accidental.

When perturbative validity is required it imposes an upper limit on
the Higgs self-coupling at our scale, and thus if Higgs bosons exist
it has been known for over a
decade\refmark{\cabibbo}that a SM Higgs boson cannot be heavier than
about 170 GeV.  The precise limit depends on $M_t$ since the top quark
Yukawa coupling enters the renormalization group equation (RGE)
for the Higgs self-coupling.

Several years ago people began to explore\REFS\habertwo{H. Haber and
M. Sher, Phys. Rev. \un{D35} (1987) 2206.}\REFSCON\gunion{J. Gunion,
L. Roszkowski and H. Haber, Phys. Lett. \un{B189} (1987) 409; see also
Phys. Rev \un{D38} (1988) 105.}\REFSCON\drees{M. Drees,
Int. J. Mod. Phys. \un{A4} (1989) 3635.}\REFSCON\binetruy{P.
Bin\'etruy and C. Savoy, Phys. Lett. \un{B277} (1992)
453.}\REFSCON\espinosa{J.R. Espinosa and M. Quiros, Phys. Lett.
\un{B279} (1992) 92.}\refsend more general supersymmetric
theories, allowing additional singlets in the Higgs sector.  A more
fundamental theory containing the MSSM could be extended through
additional Higgs multiplets (singlets, more doublets,
triplets, etc.).  Then new coupling terms could occur among these and the
two original doublets.  The effects of this
sector might have destroyed the existence of a limit if the original
Higgs fields were coupled to representations that got large vev's
which contributed to $m_\hcirc .$ It was found\refmark{\habertwo ,
\drees}that a limit still exists if any number of additional singlets
and doublets are added, once the perturbative validity condition is
imposed.  Later additional studies were done, and
recently\refmark{\espinosa}the limit was also shown to exist if Higgs
triplets were added whose vev's were kept small;
more precise numerical values were
also provided in ref. \espinosa.  Extended models could also have a larger
gauge
group or larger fermion representations, or more families could exist.
These will affect the numerical value of a limit, but not the
existence of a limit.

The present paper extends this process further, and completes it.  We
allow arbitrary Higgs representations and remove the restriction that
any vev's must remain small, allowing arbitrary triplet vev's, etc.
This is crucial for a general limit since it is known\Ref\georgi{H.
Georgi and M. Machacek, Nucl. Phys. \un{B262} (1985) 463; M.S.
Chanowitz and M. Golden, Phys. Lett. \un{B165} (1985) 105; J.F.
Gunion, R. Vega, and J. Wudka, Phys. Rev. \un{D42} (1990) 1673.}that
combinations of triplets can occur that give no contribution to the
$\rho$-parameter but have large vev's that could drive $m_\hcirc$ up.
We also add numerical contributions that could affect the value of the
limit in certain regions (e.g., the $b$ and $\tau$ contributions for
large $\tan\beta$).

It is extraordinary that the mass of the lightest Higgs boson has an
upper limit determined by weak scale parameters in a general
supersymmetric theory so long as the theory remains perturbatively
valid up to a high mass scale.
It is also remarkable that the limit is a calculable one.  It might
have happened that quantities such as soft-supersymmetry breaking
parameters that are bounded but unknown entered into the limit, e.g.
into the equations that determine the upper limits on the
self-couplings, in which case no useful numerical value could have
been obtained.

In the next section we present the derivation of the limit, and then
we present the numerical value of the limit.  Computing the precise
value of the limit is very difficult for two reasons, first because in
arbitrary extended supersymmetric theories many effects feed back on
others in ways that require extensive untangling; for example, the introduction
of new scalars increases the gauge coupling $\beta$-functions.
Second, certain properties of mass matrices are used to obtain the
upper limit, and in practice it is very hard to optimize the use of
these properties.  In this paper we present a conservative upper
limit which we hope to improve by at least several percent later.

\noindent {\bf DERIVATION}

We begin with a general superpotential and follow the same general
line as in refs. \habertwo, \drees, \espinosa.
$$W=A + B_a \varphi^a + C_{ab}\varphi^a\varphi^b+ D_{abc}
\varphi^a\varphi^b \varphi^c$$
where terms with sfermions are not written.  From this we construct
the scalar potential $V=V_F + V_D + V_{SOFT} + V_{MASS}$ in the
standard way, and separate each scalar field into real and imaginary
parts, $\varphi^a = z^{2a} + iz^{2a+1}, a=0, 1, 2, \ldots\ .$  This
gives
$$V=a+b_a z^a + c_{ab} z^az^b + d_{abc} z^a z^b z^c + f_{abcd} z^a z^b
z^c z^d + m^2_{(a)} z^a z^b \delta_{ab}$$
where the coefficients $a,b_a, c_{ab} \ldots$ can be expressed in
terms of the superpotential coefficients and the soft-breaking
coefficients.  Next the minimum of the potential is obtained by
calculating $\partial V/\partial z^i =0,$ and coefficients such as the
$m_{(a)}^2$ are eliminated using the resulting equations.  Then the
positive definite mass matrix $2M_{ij}^2 = \partial^2V/\partial z^i
\partial z^j$ is calculated, using a basis $Re\>H_1^\circ ,
Re\>H_2^\circ , Re\>S, Re\>\Sigma^\circ,\ldots$ if additional scalars
$S, \Sigma^\circ,$ etc. are present.

Finally the $2\times 2$ submatrix corresponding to $Re\>H^\circ_1,
Re\>H_2^\circ$ is examined.  The important result is that it always
has the form
$$M_{ij}^2 = \pmatrix{-J\tan\beta +K&J+L\cr J+L&-J\cot\beta +K'}.$$
Here all of the SUSY parameters and vev's that could grow are in
$J$,
$$J=J\left(m_0^2, m_{1/2}^2 , \mu, A, B, \ldots\ \hbox{vev's of new
scalars}, g_i, v_i, \lambda_i\right),$$
while $K, K', L$ depend on the gauge couplings $g_1, g_2,$ on the
vev's $v_1$ and $v_2$ that give mass to $W, Z$ (so that $v_1^2 + v_2^2
= v^2$ is fixed by $M_{\zss}$), and on the various self-couplings
$\lambda_i$.  The dependence on $\lambda_i$ means that once the
$\lambda_i$ are limited by perturbative validity the functions $K, K',
L$ have calculable upper limits. On the other hand, $J$ can become arbitrarily
large and its value is in general not calculable.

Now we observe that $TrM_{ij}^2$ and $Det\>M_{ij}^2$ both only grow as
$J$ (the $J^2$ term cancels in $Det\>M_{ij}^2$),
so one eigenvalue of $M_{ij}^2$ does not
grow with $J$ (since $TrM^2 = m_1^2 + m_2^2, Det\,M^2 = m_1^2 m_2^2$).
The eigenvalues of $M_{ij}^2$ are not the actual masses.  But the
lowest eigenvalue of an $n\times n$ positive definite matrix is less
than the smaller eigenvalue of any $2\times 2$ submatrix (imagine the
geometrical analogy of an $n$-dimensional ellipsoid and a
2-dimensional slice).  Thus the bounded eigenvalue of the $2\times 2$
submatrix is an upper limit on $m_{\hcirc}^2$.  For completeness we
give a few examples in Table 1 so the reader can see clearly how the
above argument works.

This establishes that a limit exists.  The further observation that
the RGE's for the self-couplings cannot
depend on the dimensional supersymmetry masses or soft breaking
parameters follows from general theorems, and has been exhibited
explicitly.\Ref\gato{B. Gato, J. L\'eon, J. Perez-Mercader, and M.
Quiros, Nucl. Phys. \un{B253} (1985) 285.} Thus the numerical value of
the upper limit on $m_\hcirc$ is calculable in practice.
\REF\moreespinosa{J.R. Espinosa and M. Quiros, Phys. Lett. \un{B266} (1991)
389.}

\noindent{\bf NUMERICAL VALUE OF UPPER LIMIT}

To calculate the upper limit one can proceed as follows.  The gauge
sector is fixed by $M_\zss$, and depends on $\tan\beta = v_2/v_1$.
One then can find the maximum upper limit for any $\tan\beta.$  The
upper bounds on all self-couplings are calculated from their RGE's.
Again, to be conservative we do not decide on a value $\lambda^{\max}_i$
above which a $\lambda_i$ is no longer considered to be perturbative,
but take the
value (typically a few percent larger) at which the $\lambda^{\max}_i$
saturates (the meaning of this will be clear to anyone who has worked
with such equations; we will describe the procedure in detail in a
long paper).  The $\beta$-functions for the gauge couplings $g_1$ and $g_2$
increase as more Higgs representations are added. Therefore, to calculate
the upper bound we allow the $\beta$-functions to take on their maximum
values such that the gauge couplings remain perturbative up to the high
scale.

Then we present the results as follows.  At the present time we
calculate a conservative upper limit:
\tenpoint
$$\eqalignno{&\left.\eqalign{&\hbox{Gauge and scalar
sector,}\cropen{-12pt}
&\hbox{including effects of running}\cropen{-12pt}
&\hbox{of gauge couplings,}\ t, b, \tau\cropen{-12pt}
&\hbox{Yukawas, etc.}\cr}\right\} m_{\hcirc} \leq 134\
\hbox{GeV}\left\{\eqalign{&\hbox{for}\ m_t=135\ \hbox{GeV},\cropen{-12pt}
&\hbox{decreases as}\ m_t\cropen{-12pt}
&\hbox{increases}\cr}\right.\cropen{2pt}
\noalign{\centerline{\phantom{xxxxxx}+}}\cropen{-5pt}
&\left.\eqalign{&\hbox{one-top-loop and
two-top-loop\refmark\moreespinosa}\cropen{-12pt}
&\hbox{radiative corrections}\cr}\right\}\phantom{xx} 12\
\hbox{GeV}\phantom{xx}
\left\{\eqalign{&\hbox{for}\ m_t=135\ \hbox{GeV,}\cropen{-12pt}
&\hbox{increases as}\ m_t\cropen{-12pt}
&\hbox{increases}\cr}\right.\cr}$$

\twelvepoint
\noindent We can combine them:
$$m_\hcirc \leq 146\ \hbox{GeV}\qquad\qquad\qquad\qquad
\eqalign{&\tenpoint\bigl(\hbox{\tenpoint 100 GeV\ $\leq M_t \leq 145$ GeV};
 \hbox{\tenpoint for larger $M_t$ the limit decreases,}\cropen{-8pt}
&\hbox{\tenpoint \phantom{xx}e.g. to 133 GeV at $M_t$
=}\hbox{\tenpoint 160 GeV}\bigr).\cr}$$ To this limit must be added an
amount 0-20 GeV if any new heavy fermions exist\Ref\moroi{This has
been examined by T. Moroi and Y. Okada, Mod.  Phys. Lett. \un{A7}
(1992) 187, and preprint TU-405, Tohoku Univ., Sendai, Japan}that get
their mass by the Higgs mechanism.  This number is bounded both by
perturbative validity, and by precision measurements.  A new fermion
doublet contributes to the parameter $S$ an amount $\delta
S=N_c/6\pi.$ At $2\sigma$, $S\lsim 0.37.$ As precision data improves,
the amount allowed from hypothetical new fermions will decrease.  The
heavy fermion contribution is also bounded by perturbative validity.
The precision of calculations possible so far means, we think, that
all of the above numbers are valid to at best $\pm$\ 2-3\%.  It should
be understood that the effects from top (and any other new fermion)
radiative correction loops in the Higgs effective potential $(+ 12\
\hbox{GeV and}\ + (0-20)\
\hbox{GeV})$ must be added to any upper limit, either of the MSSM or
any of its extensions.

We believe that a lengthy analysis which we are undertaking is likely
to lower the ``134 GeV'' by at least a few percent, and perhaps more.
Several effects enter, such as a present practical difficulty with
calculating the numerical value of the lower eigenvalue of the
$2\times 2$ submatrix as the parameters vary, the fact that the
parameters ($\tan\beta$, etc.) may not be able to take on
simultaneously the values that give the present conservative limit,
etc.  Also, if it were possible to set a lower limit on $\tan\beta$ of
2.5-3, the ``134 GeV'' would be lowered by over 5\%, so as the energy
of LEP increases the upper limit may decrease if no signal is found.

\noindent {\bf IMPLICATIONS}

This bound tells us that if a supersymmetric theory describes nature
on the weak scale, then a light Higgs boson can be found below the
limit.  Conversely, if no light Higgs boson is found below the limit,
no perturbative theory that requires a low energy supersymmetry to stabilize a
hierarchy of scales can be correct.  It is important to understand
that in models the mass of the lightest Higgs boson is typically
distributed from a lower value of about 70 GeV up to the
upper limit, with no strong tendency to cluster near the upper limit.  Thus
detecting the lightest Higgs boson is not likely to require a collider
that can detect $m_\hcirc = m^{\max}_\hcirc$, though possibly it
could.

In a general theory the $ZZ\hcirc$ coupling can change from its
(one-top-loop corrected) MSSM coupling.   To accompany the bound on
$m_\hcirc$ we
need to calculate the minimum cross section for $e^+e^- \to Z^{(*)}
\to Z^{(*)} + \hcirc$ over the entire parameter space, for $m_\hcirc =
m^{\max}_\hcirc$.  That is extremely difficult to do because the
singlets will not couple to the $Z$; we are currently working out what
can be said here.  For the moment we report
only that with the couplings of the radiatively corrected MSSM the
cross section will not go below $0.125pb$ for the maximum
$m_\hcirc.$

While we were writing this, additional papers studying particular
extended supersymmetry models and tightening the limits in them have
arrived.\Ref\comelli{J.R. Espinosa and M. Quiros, preprint IEM-FT-60/92;
D. Comelli and C.
Verzegnassi, DESY 92-087.}

At present the upper limit on $m_\hcirc$ is not quite within the range
where LEP could be certain to either detect $\hcirc$ or to exclude the
idea that supersymmetry is relevant to understanding nature near the
electroweak scale, even if LEP were extended to $\sqrt{s} \approx 240$
GeV.  But a combination of additional analysis of the limit (which we
are undertaking), and improved data (which will restrict the
contribution of more fermions and constrain parameters) may reduce the
upper limit significantly.  Whatever happens, LEP should be run to a
sufficiently large energy that LEP + SSC/LHC can surely detect
$\hcirc.$  Even if superpartners are directly detected, the
supersymmetric world view will not be complete until a Higgs boson is
detected.

\ack

We appreciated helpful discussions with M. Einhorn, F. Zwirner, M.
Quiros, H. Haber, E. Nardi, L. Roszkowski and R. Akhoury.
This work was supported in part by the U.S.
Department of Energy.  CK was supported in part by a University of Michigan
Regents-Crane Fellowship. JW was supported in part by a D.\/S.\/O.\/
Fellowship from the U.\/S.\/ Department of Education.
\endpage
\refout
\bye